\documentclass[11pt]{article}
\usepackage[utf8]{inputenc}
\usepackage{amsfonts,amsmath,amsxtra}
\usepackage{amssymb,amscd}
\usepackage[all]{xy}
\catcode`\@=11
\def\marginnote#1{}
\newcount\hour
\newcount\minute
\newtoks\amorpm
\hour=\time\divide\hour by60
\minute=\time{\multiply\hour by60 \global\advance\minute by-\hour}
\edef\standardtime{{\ifnum\hour<12 \global\amorpm={am}%
     \else\global\amorpm={pm}\advance\hour by-12 \fi
     \ifnum\hour=0 \hour=12 \fi
   \number\hour:\ifnum\minute<10 0\fi\number\minute\the\amorpm}}
\edef\militarytime{\number\hour:\ifnum\minute<10 0\fi\number\minute}
\def\draftlabel#1{{\@bsphack\if@filesw {\let\thepage\relax
\xdef\@gtempa{\write\@auxout{\string
   \newlabel{#1}{{\@currentlabel}{\thepage}}}}}\@gtempa
\if@nobreak \ifvmode\nobreak\fi\fi\fi\@esphack}
     \gdef\@eqnlabel{#1}}
\def\@eqnlabel{}
\def\@vacuum{}
\def\draftmarginnote#1{\marginpar{\raggedright\scriptsize\tt#1}}
\def\draft{\oddsidemargin -0.1truein
     \def\@oddfoot{\sl preliminary draft \hfil
     \rm\thepage\hfil\sl\today\quad\militarytime}
     \let\@evenfoot\@oddfoot \overfullrule 3pt
     \let\label=\draftlabel
     \let\marginnote=\draftmarginnote
\def\@eqnnum{{\rm (\theequation)}
\rlap{\kern\marginparsep\tt\@eqnlabel}%
\global\let\@eqnlabel\@vacuum}  }
\def\numberbysection{\@addtoreset{equation}{section}
     \def\theequation{\thesection.\arabic{equation}}}
\numberbysection

\renewcommand{\theequation}{\thesection.\arabic{equation}}
\makeatletter
\newdimen\normalarrayskip            
\newdimen\minarrayskip               
\normalarrayskip\baselineskip
\minarrayskip\jot
\newif\ifold             \oldtrue            \def\new{\oldfalse}
\def\arraymode{\ifold\relax\else\displaystyle\fi}
\def\eqnumphantom{\phantom{(\theequation)}} 
\def\@arrayskip{\ifold\baselineskip\z@\lineskip\z@
  \else
  \baselineskip\minarrayskip\lineskip1\baselineskip\fi}
\def\@arrayclassz{\ifcase \@lastchclass \@acolampacol \or
\@ampacol \or \or \or \@addamp \or
\@acolampacol \or \@firstampfalse \@acol \fi
\edef\@preamble{\@preamble
\ifcase \@chnum
  \hfil$\relax\arraymode\@sharp$\hfil
  \or $\relax\arraymode\@sharp$\hfil
  \or \hfil$\relax\arraymode\@sharp$\fi}}
\def\@array[#1]#2{\setbox\@arstrutbox=\hbox{\vrule
  height\arraystretch \ht\strutbox
  depth\arraystretch \dp\strutbox
width\z@}\@mkpream{#2}\edef\@preamble{\halign \noexpand\@halignto
\bgroup \tabskip\z@ \@arstrut \@preamble \tabskip\z@ \cr}%
\let\@startpbox\@@startpbox \let\@endpbox\@@endpbox
\if #1t\vtop \else \if#1b\vbox \else \vcenter \fi\fi
\bgroup \let\par\relax
\let\@sharp##\let\protect\relax
\@arrayskip\@preamble}
\def\eqnarray{\stepcounter{equation}%
           \let\@currentlabel=\theequation
           \global\@eqnswtrue
           \global\@eqcnt\z@
           \tabskip\@centering              
           \let\\=\@eqncr
           $$%
         \halign to \displaywidth  \bgroup
          \eqnumphantom \@eqnsel
   \hskip\@centering                               
 $\displaystyle  \tabskip\z@ {##}$%
 &\global\@eqcnt\@ne \hskip 2\arraycolsep
      $ \displaystyle  \arraymode{##}$\hfil
 &\global\@eqcnt\tw@ \hskip 2\arraycolsep
      $\displaystyle\tabskip\z@{##}$\hfil
      \tabskip\@centering
 &{##}\tabskip\z@\cr}
\makeatother

\newcounter{mo}

\newcounter{bk}

\newcommand{\Si}{\Sigma}
\newcommand{\tr}{{\rm tr}}

\newcommand{\ti}[1]{\tilde{#1}}
\newcommand{\om}{\omega}

\newcommand{\de}{\delta}
\newcommand{\al}{\alpha}

\newcommand{\vth}{\vartheta}
\newcommand{\be}{\beta}
\newcommand{\la}{\lambda}
\newcommand{\La}{\Lambda}

\newcommand{\G}{\Gamma}
\newcommand{\ka}{\kappa}

\newcommand{\ga}{\gamma}

\def\bea{\begin{eqnarray}\new\begin{array}{cc}}
\def\ee{\end{array}\end{eqnarray}}

\newcommand{\beq}[1]{\begin{equation}\label{#1}}
\newcommand{\eq}{\end{equation}}
\newcommand{\beqn}[1]{\begin{small} \begin{eqnarray}\label{#1}}
\newcommand{\eqn}{\end{eqnarray} \end{small}}
\newcommand{\p}{\partial}
\def\sq2{\sqrt{2}}
\newcommand{\di}{{\rm diag}}
\newcommand{\oh}{\frac{1}{2}}

\newcommand{\GLN}{{\rm GL}(N,{\mathbb C})}
\def\sln{{\rm sl}(N, {\mathbb C})}
\def\sl2{{\rm sl}(2, {\mathbb C})}
\def\SLN{{\rm SL}(N, {\mathbb C})}

\newcommand{\PGLN}{{\rm PGL}(N,{\mathbb C})}
\newcommand{\gln}{{\rm gl}(N, {\mathbb C})}

\newcommand{\SON}{{\rm SO}(N, {\mathbb C})}
\newcommand{\son}{{\rm so}(N, {\mathbb C})}

\def\f1#1{\frac{1}{#1}}

\def\mC{{\mathbb C}}
\def\mZ{{\mathbb Z}}

\def\frak{\mathfrak}

\def\gg{{\frak g}}
\def\gp{{\frak p}}

\def\gk{{\frak k}}

\def\gh{{\frak h}}

\def\bfe{{\bf e}}

\def\bfu{{\bf u}}
\def\bfv{{\bf v}}

\def\bfC{{\bf C}}

\def\bfS{{\bf S}}

\def\bfX{{\bf X}}

\def\bfW{{\bf W}}

\def\clR{\mathcal{R}}

\def\clO{\mathcal{O}}

\def\clM{\mathcal{M}}

\def\clP{\mathcal{P}}
\def\clQ{\mathcal{Q}}

\def\clX{\mathcal{X}}

\def\clZ{\mathcal{Z}}

\def\bag2{{\bf g_2}}
\def\bas8{{\bf so(8)}}

\def\sr2{\sqrt{2}}

\def\f1#1{\frac{1}{#1}}

\def\rank{{\rm rank}}

\begin{document}

 \begin{flushright}
 ITEP-TH-8/21\\
 IITP-TH-5/21
 \end{flushright}
\vspace{3mm}

 \begin{center}
{\LARGE Integrable extensions of classical elliptic integrable systems}\\
 \vspace{10mm}

  {\large M. Olshanetsky}\\
   \vspace{7mm}
   {\texttt NRC Kurchatov Institute, ITEP, Moscow, B. Cheremushkinskaya, 25, Moscow, 117259, Russia}\\
   {\texttt Institute for Information Transmission Problems RAS (Kharkevich Institute),
 \\  Bolshoy Karetny per. 19, Moscow, 127994,  Russia}\\
{\texttt Moscow Institute of Physics and Technology,\\
 Inststitutskii per.  9,  Dolgoprudny, 141700 Moscow Region, Russia}\\
\vspace{4mm}
 {\footnotesize Email olshanet@itep.ru}\\
 \end{center}

 \begin{flushleft}
\emph{In Memory of Mikhail Konstantinovich Polivanov}
\end{flushleft}

 \begin{abstract}
 In this article we consider two particular examples of general
 construction proposed in arXiv:2012.15529.
We consider the integrable extensions of the classical elliptic Calogero-Moser model
of N particles  with spin and
the integrable Euler-Arnold top related to the group SL(N,C). The
extended systems has additional N-1 degrees of freedom and can be described in terms of the Darboux
variables.

\end{abstract}



\section{Introduction and summary}
\setcounter{equation}{0}

It is possible to extend the phase space of the integrable
such that an integrable system exists on the extended phase space. The extension is non-trivial if the extended system is not a direct product of two non-interacting systems.
We have not a general theory of such extensions, and consider some examples based on the extensions
of Hitchin type systems \cite{Hi,Ne}. In our construction
the underlying system is the symplectic quotient of the extended system with respect to the action
of some abelian group.

A particular example of a non-trivial extension that does not, however, fit into our scheme
is the Hitchin systems related to parabolic Higgs bundles (the Hitchin systems with spin variables). The symplectic quotients of these systems are the Hitchin systems defined on  smooth curves (the systems without spin.

An example of such a non-trivial expansion is the passage from the system of $N$ Calogero-Moser (CM) particles, to the same system of particles equipped  with spin \cite{GH,Wo}.
Here, in particular,  we consider the further extension of this  system to the system with additional
$N-1$ degrees of freedom.

 In our previous publication \cite{LOZ1} we proposed some generalizations
 of the parabolic Higgs bundles which lead to a special extensions of integrable systems.
 The advantage of our construction is the existence of Darboux coordinates on the
 extended system.

 The first example of these systems is the Sutherland model two types of spins
 \cite{FP,KLOZ}.
Here we consider in details two additional examples. The first one is the extension of
integrable Euler-Arnold top related to the group $\SLN$ \cite{RS,KLO}.
The second is the extension of spin elliptic CM system.

The next chapter is auxiliary.
We describe the symplectic structure on one of the symmetric $\SLN$-spaces and on the $\SLN$-coorbits  and establish the relation between them.
In the next chapter we actually consider the above two examples

\section{Cotangent bundles to symmetric spaces and coadjoint orbits}
\setcounter{equation}{0}

\subsection{Cotangent bundles to groups.}
Consider the complex groups $\GLN$, $\SLN$ and P$\GLN$ for $N\geq 2$.
$$
\GLN=\{f\in Mat_N\,|\,\det\,f\neq 0\}\,,
$$
\beq{d1}
\SLN=\{f\in\GLN\,|\,\det\,f=1\}\,,~~\SLN\subset\GLN\,,
\eq
$$
\PGLN=\GLN/\mC^*\,,
$$
where $\mC^*$ is the center of $\GLN$.
Let $\clZ=\mZ_N$ be the center of $\SLN$. Then
\beq{sp}
\PGLN=\SLN/\clZ\,.
\eq

Let $G$ be the group $\GLN$, $\SLN$ or $\PGLN$ and $\gg$ their  Lie algebras.
We identify $\gg$ with the Lie coalgebra $\gg^*$ by means the invariant metric $(~,~)$
on $\gg$. Then the algebra $\gg$ becomes a Poisson space.
Let  $\bfS=\sum S^aT_a$ be the expansion in a basis of $\gg$.
The Poisson brackets  are the Lie-Poisson brackets
\beq{lb}
\{S^a,S^b\}=C^{ab}_cS^c\,,
\eq
where $C^{ab}_c$ are structure constants of the algebra $\gg$.
These brackets are degenerated on  $\gg$. Fixing $l=\rank\,\gg$ Casimir functions
\beq{fc1}
(\bfS)=c_1\,,\,(\bfS^2)=c_2\,,\ldots, (\bfS^{N})=c_{N}
\eq
we come to the non-degenerated brackets on co-adjoint orbits $\clO_\nu$ described below.
The dimension of the generic orbit is $\dim\,(\clO_\nu)=\dim\,G-l$.
In what follows we will consider only these types of orbits.

The cotangent bundle $T^*G$ is identified with the tangent bundle $T\,G$. Using its trivialization
 we describe $T^*G$ as
 \beq{cob}
 T^*G=\gg\times G=\{\zeta\in \gg\,,\,g\in G\}\,.
 \eq
 The canonical invariant symplectic form on $T^*G$
in these coordinates takes the form
  \beq{sfcb}
\om=D(\zeta,Dg g^{-1})=(\bfX,g^{-1}Dg)\,,
\eq
where
\beq{bx}
\bfX=g^{-1}\zeta g\in\gg\,.
\eq

The form is invariant under the left action of subgroup $K\subseteq G$
  \beq{ag1}
\zeta\to f\zeta f^{-1}\,,~~g\to fg\,,~~\bfX\to\bfX\,,~~f\in K\,.
  \eq
producing the left moment
\beq{rm1}
\mu^L(\zeta,g)=\zeta|_{\gk^*}=g\bfX g^{-1}|_{\gk^*}\,,
\eq
where $\gk=$Lie$(K)$, and $\gk^*$ is the Lie coalgebra.

The right action
\beq{ag5}
\zeta\to\zeta \,,~~g\to gf^{-1}\,,~~f\in G
\eq
leads to the moment map
\beq{lm1}
\mu^R(\zeta,g)=g^{-1}\zeta g=\bfX\,.
\eq

Consider the interrelations between $T^*\GLN$ and $T^*\SLN$.
The symplectic form $\om$ (\ref{sfcb}) on $T^*\GLN$ is invariant under the scaling
$$
g\to g\la \,, ~~\zeta\to\zeta\,,~~\la\in\mC^*\,.
$$
The corresponding moment is $\tr\,\zeta$. We fix the gauge by the condition $\det\,g=1$ (\ref{d1}).
In this way we come to $T^*\SLN$ as the symplectic quotient
\beq{sq1}
T^*\SLN=\mC^*\setminus\!\setminus T^*\GLN=\{(\zeta,g)\,|\,\tr\,\zeta=0\,,~\det\,g=1\}\,.
\eq
The condition $\det\,g=1$ does not fix the gauge completely. One can further act by the center
$\clZ$ on $g\to\ga g$ that preserves this condition. In this way we come to the cotangent bundle
$T^*\PGLN$.

The symplectic form on $T^*\SLN$ is
\beq{sf1}
\om=D( \zeta,D g g^{-1})=(\bfX,g^{-1}Dg)\,,~~{\rm where}~(\zeta,g)~{\rm satisfy}~(\ref{sq1})\,.
  \eq
Since $\bf X$ is gauge invariant and $\tr\,\bfX=0$ it is a section of
the cotangent bundle $T^*\SLN$.

There is a residual symmetry defined by multiplication on the center element
\beq{ce}
\bfe_N(\ga)\,,~\ga=\di(1,\ldots,1)\,,~~\bfe_N(\ga)=\exp\,\frac{2\pi\imath\ga}N\,,~
\bfe_N(\ga)\in\clZ\,.
\eq
After factorizing we come to the cotangent bundle $T^*\PGLN$.


\subsection{Co-adjoint orbits}
Consider the cotangent bundle  $T^*\SLN$.
 and the left symplectic quotient\\
  $\SLN\setminus\setminus_\nu T^*\SLN$.
with the moment $\mu^L$ (\ref{rm1}) taking value in the Cartan subalgebra $\gh^\mC\subset\sln$
\beq{or}
\mu^L(\zeta,g)=\zeta =\nu\in (\gh^\mC)^*\sim\gh^\mC\,.
\eq
Here $\nu$ is a fixed regular element of $\gh^{\mC}$. The subgroup preserving this value is the Cartan subgroup $H^\mC\subset\SLN$.
Thus, the symplectic quotient is defined as
\beq{or2}
\SLN\setminus\!\setminus_\nu T^*\SLN=\{(\zeta,g)\,|\, \zeta=\nu\in\gh^\mC\,,~g\sim fg\,,\,f\in H^\mC\}\,.
\eq
It is the co-adjoint orbit $\clO_\nu$
\beq{orb}
\clO_\nu=H^\mC\setminus \SLN
=\{\bfS=g^{-1}\nu g\,|\,g\in \SLN,,~\nu\in\gh^\mC\}\,.
\eq

After substituting $\zeta=\nu$ (\ref{or}) in (\ref{sfcb}) the form $\om$ on $T^*\SLN$ becomes the  Kirillov-Kostant on $\clO_\nu$
form
  \beq{kkf}
\om^{KK}=D(\nu,D g g^{-1})\,.
  \eq
  The form is invariant under the right $G^\mC$ action (\ref{ag1}).
  This transformation generates the moment
  \beq{rmo}
  \mu^R=g^{-1}\nu g=\bfS\,.
  \eq

For $\gg=\sln$ the Casimir functions (\ref{fc1}) are
\beq{fc}
(\bfS^2)=c_2\,,\ldots, (\bfS^{N})=c_{N}\,,~~
(c_2=\sum_{j=1}^N\nu^2_j\,,\ldots,c_N=\sum_{j=1}^N\nu^N_j)
\eq
and we assume that $c_j\neq c_k$ for $j\neq k$. The corresponding orbit is
\beq{co}
\clO_\nu=g^{-1}\nu g\,,~g\in\SLN\,,~ \nu=\di(\nu_1,\ldots,\nu_N)\,,~\tr\,\nu=0\,,
\eq
and $\nu_j\neq \nu_k$ for $j\neq k$.
The dimension of the generic orbit is
   \beq{dor}
 \dim_\mC\,(\clO_\nu)=\dim_\mC\,\SLN-(N-1)=N(N-1)\,.
     \eq


\subsection{Cotangent bundles to $T^*(\SON\setminus\GLN$) }

Consider the group $G=\GLN$, its subgroup
\beq{o}
K=\SON=\{g\in\GLN\,|\,g^Tg=Id\,,~\det\,g=1 \}
\eq
and the quotient space $\clX^G=\SON\setminus\GLN$.
Thereby $\clX^G=\{x\}$ is the space of complex symmetric matrices $x$ with $\det\,x\neq 0$.
Evidently, the element
\beq{q1}
\clQ=g^Tg\,,~~g\in\GLN
\eq
is an element of $\clX^G$.

The cotangent bundle
 $T^*\clX^G=T^*(\SON\setminus\GLN)$ can be identified with
 the symplectic quotient $\SON\setminus\!\setminus T^*G$.
The form $\om$ (\ref{sfcb}) on $T^*\GLN$ is invariant under the left $\SON$-action
\beq{ag2}
\zeta\to f\zeta f^{-1} \,,~~g\to fg\,,~~f\in \SON\,.
  \eq
The corresponding moment map assumes the form $\mu^L=\zeta |_{\son}$ and we take
\beq{mcc}
\zeta |_{\son}=0\,.
\eq
Note that  the $\SON$-action preserves the moment constraint $f\zeta f^{-1}|_{\son}=0$ for $f\in \SON$.

The Lie algebra $\gg=\gln$ can be decomposed in the sum of the antisymmetric and symmetric matrices
\beq{cd}
\gln=\son+\gp^G\,.
\eq
The space of symmetric matrices $\gp^g$ can be considered as
 the tangent space to $\clX^G=\{\SON g\}$ at the point $g=Id$. These subspaces are orthogonal with respect to the invariant metric on $\gln$. Its restriction  on  $\gp^G$ is non-degenerate.

The moment condition (\ref{mcc}) means
that $\zeta\in\gp^G$.
In this way the symplectic quotient $\SON\setminus\!\setminus T^*\GLN$
 is defined by the set of pairs $(g,\zeta)$ that
satisfy the equivalence relation
\beq{pv}
\{(g,\zeta)\sim(fg, f\zeta f^{-1})\,,~~ f\in \SON\,,~g\in\GLN\,~~\zeta\in\gp^G\}\,.
\eq
Since $g\in\GLN$ is defined up to the left multiplication by $\SON$, we come to the cotangent bundle
$T^*\clX^G$, $\clX^G=\SON\setminus\GLN$
\beq{txg}
T^*\clX^G=\SON\setminus\!\setminus T^*\GLN\,.
\eq

 It follows from this definition of $T^*\clX^G$ that
\beq{dcb}
\dim\,T^*\clX^G=2(\dim\,\GLN-\dim\,\SON)=N(N+1)\,.
\eq

The symplectic form on $T^*\clX^G$ coincides with (\ref{sfcb})
\beq{sfq}
\om=D(\zeta,Dg g^{-1})\,,
\eq
where the pair $(\zeta,g)$ satisfies (\ref{pv}).

Let $g\in\GLN$ and $\zeta\in\gp^G$.
Consider the element
\beq{xr}
\bfX^G=g^{-1}\zeta g\in\gln\,.
\eq
It is a section of the bundle  $T^*\clX^G$.
In this terms $\om$ (\ref{sfq}) assumes the form
\beq{sfq1}
\om=D(\bfX^G,g^{-1}Dg )\,.
\eq
 Since $\bfX^G\in\gln$ it can be expanded in the basis $T_a$
of the algebra $\gln$\\
 $\bfX^G=\sum X^aT_a$. The coefficients
$X^a$ form the Lie-Poisson algebra on $\gln$. As in (\ref{lb})
\beq{lb1}
\{X^a,X^b\}=C^{ab}_cX^c\,,~~(a,b=1,\ldots,N^2)\,.
\eq
As above, these brackets have $N$ Casimir functions $c_j=(\bfX^{j})$, $\,(j=1,\ldots,N)$.
But they are not the Casimir functions of the Poisson algebra on $T^*\clX^G$.

 It follows from (\ref{ag2}) that
  $T^*\clX^G$ can be described by the gauge-invariant variables
\beq{cav}
\clP=g^{-1}\zeta(g^T)^{-1}\,,~~\clQ=g^Tg\,,~(\ref{q1})
\eq
where $(\zeta,g)$ satisfies (\ref{pv}).
Note that $\clP^T=\clP$ and $\clQ^T=\clQ$ ($\clQ\in\clX^G$). In these variables
\beq{xr1}
\bfX^G=\clP\clQ\,.
\eq

 Let $e_j$  be a basis in  $\gp^G$ $(j=1,\ldots,\dim\,\gp^G)$ with the pairing $(e_j,e_k)=\de_{jk}$.
 Expand $\clQ$  and $\clP$ in this basis $\clQ=\sum_j\clQ^je_j$, $\clP=\sum_j\clP^je_j$.
 In these variables the Poisson algebra on $T^*\clX$ is canonical
 \beq{pb2}
 \{\clP^j,\clQ^k\}=\de^{jk}\,,~~(j,k=1,\ldots,\dim\,\gk)\,.
 \eq
 To prove it define another invariant symplectic form on $T^*\clX^G$
 \beq{ox}
\om^{X^G}=\om(\zeta,g)-\om(\zeta,(g^T)^{-1})\,,
\eq
where $\om$ is (\ref{sfq}). By direct calculations we find that
\beq{caf1}
\om^{X^G}=(D\clP,D\clQ)\,.
\eq
In this way we come to the Darboux brackets (\ref{pb2}).

 The symplectic form $\om$ (\ref{sfq1}) on $T^*\clX^G$ as well $\om^X$ (\ref{caf1})
 is invariant under the right action of the group $G$
(\ref{ag5}). Similarly to (\ref{rm1}) the moment corresponding to the action is
\beq{rm0}
\mu^R(\zeta,g)=g^{-1}\zeta g\in\gg^\mC|_{Lie^*(G)}\,,
\eq
(see (\ref{mcc})),  or
  \beq{mla1}
  \mu^R\stackrel{(\ref{xr})}=\bfX^G\stackrel{(\ref{xr1})}=\clP\clQ\,.
  \eq


\subsection{Cotangent bundle $T^*(\SON\setminus\SLN)$ and coadjoint orbits}
In what follows we need the quotient space $\clX^S=\SON\setminus\SLN$.
 It is the space of complex symmetric matrices with $\det\,g=1$
 \beq{xs1}
 \clX^S=\{g\in\SLN\,|\,g^T=g\}\,.
 \eq
  In fact,
 it  is a symmetric pseudo-riemannian
space (see definitions in \cite{Be,He}).

Similarly to (\ref{pv}) the cotangent bundle $T^*\clX^S$
is the result of the symplectic reduction of the
cotangent bundle $T^*\SLN$ under the left action of the subgroup SO$(N,\mC)$
$$
T^*\clX^S =T^*(\SON\setminus\SLN)=\SON\setminus\!\setminus T^*\SLN\,.
$$
Let $\gp^S$ be the subspace in the Lie algebra $\sln$ orthogonal to the
Lie subalgebra $\son$
$$
\sln=\son+\gp^S\,,~~
\gp^S=\{\zeta\in\sln\,|\,\zeta^T=\zeta\,,~\tr\zeta=0\}\,.
$$
In these terms the cotangent bundle $T^*\clX^S$ can be identified with the set of pairs
\beq{ts}
T^*\clX^S=\{(g,\zeta)\,,\,\,g\in\clX^S\,(\ref{xs1})\,,\,\zeta \in\gp^s\}\,.
\eq
In this way
the symplectic  form $\om$ on $T^*\clX^S$ is
\beq{os}
\om^S=D(\zeta,Dgg^{-1})=D(\bfX^S,g^{-1}Dg)\,,~~g\in\clX^S\,,~\bfX=g^{-1}\zeta g\,.
\eq
It follows from (\ref{dcb}) that
\beq{dcb1}
\dim\,T^*\clX^S=N(N+1)-2=\dim\,\clO_\nu+2(N-1)\,.
\eq
In terms of Darboux variables  $\bfX^S=\clP\clQ$ (\ref{xr}),
where $\clP$ and $\clQ$ are complex symmetric matrices with $\det\,g=1$ and $\tr\,\bfX^S=0$.

Consider the symplectic action of $\mC^*$ on $T^*\clX^G$
$$
(g,\zeta)\to (\la g,\zeta)\,.
$$
Since this action commutes with the $\SON$-action we have  as in (\ref{sq1}) another realization of
$T^*\clX^S$
\beq{xs}
T^*\clX^S=\mC^*\setminus\!\setminus T^*\clX^G\,.
\eq

 The Darboux variables have the form (\ref{cav})
\beq{pq1}
\clP=g^{-1}\zeta(g^T)^{-1}\,,~~\clQ=g^Tg\,,\,,~~\det\,g=1\,,~\tr\,\zeta=0
\eq
the form is
$$
\om^{X^S}=\om^S(\zeta,g)-\om^S(\zeta,(g^T)^{-1})\,,
$$
where $\om^S$ is (\ref{os}) and
\beq{caf2}
\om^{X^S}=(D\clP,D\clQ)\,.
\eq
The additional constraints in these terms are
\beq{sc}
1.\,\tr\,\bfX^S=\tr\,\clP\clQ=0\,,~~2.\,\det\,\clQ=1\,.
\eq

 Since the Cartan subgroup
$H^\mC\subset\SLN$ lies in $\clX^S$
($H^\mC\notin$SO$(N,\mC)$)
 one can consider the additional left action of  $H^\mC$ on $T^*\clX^S=\SLN\setminus\setminus_\nu\SLN$
\beq{lda}
g\to fg\,,~~\zeta\to f\zeta f^{-1}\,, ~~f\in H^\mC\,.
\eq

Let $\gh^\mC\subset\sln$ be the Cartan subalgebra $\gh^\mC=$Lie$(H^\mC)$ and $\nu\in\gh^\mC$
 is a regular element. The symplectic reduction of $T^*\clX^S$ with respect to this action is defined by the moment constraint equation
\beq{pao}
\mu^L=Pr\zeta \,|_{\gh^\mC}=\nu
\eq
and the gauge invariant variable $\bfX^S$ (\ref{xr}).
We take $\nu$ to be the same as in (\ref{or2}). Therefore $\bfX^S=g^{-1}\nu g\in\gg^\mC$
is an element of the coadjoint orbit $\clO_\nu$ (\ref{orb}). It means that
\beq{oga}
\clO_\nu=\{\bfX=g^{-1}\nu g\,|\,g\in G^\mC\}=H^\mC\setminus\setminus_\nu T^*\clX^S\,.
\eq

Summarizing we obtain the   descriptions of symplectic manifolds incorporated in
 the commutative diagram, where the arrows mean
the symplectic reductions
$$
\xymatrix{
  \ar[d]_{2.} T^*\GLN \ar[r]^{1. }
  & T^*\SLN
   \ar[d]_{3. }
   \ar@/^3pc/[dd]^{6.}\\
      T^*\clX^G \ar[r]_{4.}
         & \quad  T^*\clX^S
       \ar[d]_{5.} \\
  & \clO_\nu
              }
$$

$$
 \begin{tabular}{|c|c|c|c|}
  \hline
  \hline
  &Simplectic action   & Constraints &Definition \\
  \hline
  \hline
  1.& $\mC^*\setminus \!\setminus$ & $\det\,g=1\,,~\tr\,\zeta=0$ &(\ref{sq1})\\
  2.& $\SON\setminus \!\setminus$ & $g^T=g\,,~\zeta^T=\zeta$ &(\ref{txg})  \\
 3.&  $\SON\setminus \!\setminus$ & $g^T=g\,,~\zeta^T=\zeta$ &(\ref{ts})  \\
  4.&  $\mC^*\setminus \!\setminus$& $\det\,g=1\,,~\tr\,\zeta=0$ &(\ref{xs})\\
   5.&  $H^\mC\setminus \!\setminus_\nu$& $g\in\SLN\,,~\zeta=\nu\in\gh^\mC$ &(\ref{oga})\\
    6.&  $\SLN\setminus \!\setminus_\nu$& $g\in\clX^S\,,~\zeta=\nu\in\gh^\mC$ &(\ref{orb})\\
 \hline
   \end{tabular}
$$
\begin{center}
   \texttt{Table}
   \end{center}




\section{Extensions of elliptic integrable systems}
\setcounter{equation}{0}

%
%

Let $\Si_\tau=\mC/(\mZ+\tau\mZ)$ be the elliptic curve.
The Lax operator $L(z)$ is a meromorphic $(1,0)$-form on $\Si_\tau$,
taking values in the Lie algebra $\sln$.
It satisfies some quasi-periodicity conditions with respect to the shifts
on the lattice vectors and has a simple pole at $z=0$.
The residue and the quasi-periodicities fix $L(z)$.
 Let
 \beq{rf}
Res\,L(z)|_{z=0}=\bfW\,.
\eq
We consider two types of the residues
\beq{ca}
\bfW=\left\{
\begin{array}{ll}
  1. & \bfS\in\clO_\nu\,, \\
  2. & \bfX\in T^*\clX^G~{\rm or}~ T^*\clX^S
\end{array}
\right.
\eq
The first case corresponds to standard integrable systems.
The formula (\ref{dcb1}) suggests that the Lax operator in the
latter case defines the extension of the standard integrable systems.
We will prove that this extension is also integrable.

We consider  two types of the quasi-periodicities.
The first one defines the integrable Euler-Arnold $\SLN$ top and the second
the elliptic CM system with spin.

\subsection{Extension of  integrable Euler-Arnold $\SLN$ top}

\noindent
{\bf Euler-Arnold $\SLN$ top}\\
To define the Euler-Arnold top introduce a special basis in the groups $\GLN$, $\SLN$ and the corresponding Lie algebras.
Consider  two matrices
\beq{q}
Q=\di({\bf e}_N(1),\ldots,{\bf e}_N(m),\ldots,1) \,,~~{\bf e}_N(z)=\exp (\frac{2\pi i}{N} z)
\eq
\beq{la} \La= \left(\begin{array}{ccccc}
0&1&0&\cdots&0\\
0&0&1&\cdots&0\\
\vdots&\vdots&\ddots&\ddots&\vdots\\
0&0&0&\cdots&1\\
1&0&0&\cdots&0
\end{array}\right)\,.
\eq
 Note that $Q^N=\La^N=Id$ and
 \beq{ql}
Q^m\La^n=\bfe_N(-mn)\La^nQ^m\,.
\eq

 Consider two-dimensional lattices in $\mC$
 \beq{B.10}
\Gamma_N=\mZ^{(2)}_N=(\mZ/N\mZ\oplus\mZ/N\mZ)\,,~~\ti{\Gamma}_N=\ti{\mZ}^{(2)}_N=
\mZ^{(2)}_N\setminus(0,0)
 \eq

  The matrices
$Q^{a_1}\La^{a_2}$, $a=(a_1,a_2)\in\mZ^{(2)}_N$ generate a basis in
the group $\GLN$ and the algebra $\gln$, while $Q^{\al_1}\La^{\al_2}$,
$\al=(\al_1,\al_2)\in\ti{\mZ}^{(2)}_N$ generate a basis in the Lie
algebra $\sln$.
More exactly, define the generators
\beq{B.11}
  T_{a}= \frac{N}{2\pi\imath}\bfe_N(\frac{a_1a_2}{2})Q^{a_1}\La^{a_2}\,.
 \eq
 They are almost invariant on the lattice $\mZ\oplus\mZ$:
 $T_{a_1+N,a_2+N}=\pm T_{a_1,a_2}$. Therefore, $T_{- a_1,- a_2}=\pm T_{N-a_1,N-a_2}$.

From (\ref{ql}) one finds the multiplication in $\GLN$
 \beq{AA3a}
T_aT_b=\ka_{a,b}T_{a+b}\,,\ \ \ka_{a,b}=\frac{N}{2\pi
i}\bfe_N(-\frac{a\times b}{2}),\ ~~ (a\times b=a_1b_2-a_2b_1)\,.
 \eq
Similarly, the commutation relations for the algebra $\sln$ in the basis
 $T_\al$, $\al\in\ti{\Gamma}_N$ assume the form
\beq{AA3b}
[T_{\al},T_{\be}]=\bfC(\al,\be)T_{\al+\be}\,,
~~\bfC(\al,\be)=\frac{N}{\pi}\sin\frac{\pi}{N}(\al\times \be)
\eq

Define invariant form on $\gln$
\beq{ifo}
 (T_a,T_b)=\frac{4\pi^2}{N^2}\tr\,(T_a\cdot T_b)=\de_{a+b,0}\,.
\eq

Consider the expansion of the elements of an element $\bfS\in\sln$
$$
\bfS=\sum_{\al\in\ti{\Gamma}_N}S_\al T_\al\,.
$$
From (\ref{AA3b})
the corresponding Poisson brackets are
\beq{pb1}
\{S_{\al},S_{\be}\}=\bfC(\al,\be)S_{\al+\be}\,.
\eq
Fix the Casimir functions (\ref{fc}). It means that $\bfS$ belongs to the corresponding
coadjoint $\clO_\nu$ (\ref{orb}).

Let $\wp(x)$ be the Weierstrass function (\ref{wp}) and
$\wp_\al=\wp(\frac{\al_1+\al_2\tau}N)$, ($\al\in\ti{\Gamma}_N$).
The integrable Euler-Arnold $\SLN$-top is defined by the Hamiltonian
\beq{ha1}
H=\oh\sum_{\al\in\ti{\Gamma}_N}S_\al\wp_\al S_{-\al}
\eq
and the Lie-Poisson brackets (\ref{pb1}).
The phase space $\clM^{ET}$ of the elliptic top is the coadjoint orbit
\beq{pst}
\clM^{ET}\sim\clO_\nu\,.
\eq
 To find the commuting integrals
of motion we use the Lax operator.


  \bigskip
\noindent
\emph{\textbf{Lax operator }}\\
In this case  the Lax operator $L(z)$ has the quasi-periodicities
\beq{lqp}
L^{ET}(z+1)=QL^{ET}(z)Q^{-1}\,,~~L^{ET}(z+\tau)=\La L^{ET}(z)\La^{-1}\,.
\eq

It can be defined in terms of the Kronecker function.
The Kronecker function $\phi(u,z)$ is related to the elliptic curve $\Si_\tau$
and takes the form
\beq{phi}
 \phi(u,z)=\frac{\vth(u+z)\vth'(0)}{\vth(u)\vth(z)}\,,
 \eq
 where $\vth(z)$ is  the theta-function
\beq{theta}
\vth(z|\tau)=q^{\frac
{1}{8}}\sum_{n\in {\bf Z}}(-1)^ne^{\pi i(n(n+1)\tau+2nz)}\,.
\eq
In addition we need the Eisenstein function
$$
E_1(z)=\p_z\log\vth(z)\,.
$$
The $\eta_1$ constant is extracted from the asymptotic of $E_1(z)$
$$
E_1(z)\sim\f1{z}+\eta_1z+\ldots\,.
$$
The Weierstrass function $\wp$ is related to $E_1(z)$ as
\beq{wp}
\wp(u)=-\p_zE_1(u)-2\eta_1
\eq
is double-periodic function with a second order pole at $z=0$.

The basis of the doubly periodic functions on $\Si_\tau$ are
\beq{dp}
\{1\,,\,(-\p_z)^{k-2}\wp(z)\,,k=2,3,\ldots\}\,.
\eq
They have poles of orders $0,2,3,\ldots$.

The Kronecker function has the following  quasi-periodicities:
\beq{A.14}
\phi(u,z+1)=\phi(u,z)\,,~~~\phi(u,z+\tau)=e^{-2\pi \imath u}\phi(u,z)\,.
\eq
and has the first order pole at $z=0$
\beq{rk}
\phi(u,z)\sim\f1{z}+E_1(u)+\oh z(E^2_1(u)-\wp(u))+\ldots\,.
\eq
It is related to the Weierstrass function $\wp$ as follows:
\beq{wpphi}
\phi(u,z)\phi(-u,z)=\wp(z)-\wp(u)\,.
\eq
Let
$$
\varphi_{m}(z)=\bfe_N(-m_2 z)\phi(-\frac{m_1+m_2\tau}N,z)\,, ~~(m=(m_1,m_2))\,.
$$
Then
\beq{vf}
\varphi_{m}(z+1)=\bfe_N(-m_2)\varphi_{m}(z)\,,~~
\varphi_{m}(z+\tau)=\bfe_N(m_1)\varphi_{m}(z)\,.
\eq
The Lax operator assumes the form
\beq{l1}
L^{ET}(z)=\sum_{\al\in\ti\G_N}S_\al\varphi_{\al}(z)T_\al\,.
\eq
It follows from (\ref{A.14}), (\ref{AA3a}) that $L(z)$ has needed quasi-periodicities (\ref{lqp}).
Note, that there are no constant terms because it does not satisfy (\ref{lqp}).
From (\ref{rk})
\beq{re}
Res\,L^{ET}(z)|_{z=0}=\sum_{\al\in\ti\G_N}S_\al T_\al\,.
\eq

These properties imply that $(L(z)^k)$ are doubly periodic functions with the poles up to the
order k. Thereby, they can be expanded in the basis of the Weierstrass function and its derivatives
(\ref{dp})
\beq{lk}
((L^{ET})^k(z))=I_{0,k}+I_{2,k}\wp(z)+\ldots+I_{k,k}(\p_z)^{k-2}\wp(z)\,.
\eq
The term $I_{1,k}$ is absent because due to the residue theorem there are no periodic
functions with one simple pole on $\Si_\tau$.
In particular, from (\ref{wpphi})
\beq{qh}
((L^{ET})^2(z))=I_{0,2}+I_{2,2}\wp(z)\,,~~I_{0,2}=-2H^{ET}\,,~
(\ref{ha1})\,.
\eq

Thus, there are $\oh N(N+1)-1$ independent quantities $I_{s,k}$ ($k=2,3,\ldots,N$, $s=0,2,\ldots,k$)
\beq{ti}
\begin{array}{ccccc}
I_{0,n}&I_{2,N}&\dots&\dots&I_{N,N}\\
\dots&\dots&\dots&\dots&\dots\\
&I_{0,3}&I_{2,3}&I_{3,3}&  \\
\end{array}
\eq
\vspace{-0.3cm}
$$
\begin{array}{cc}
I_{0,2}&I_{2,2}\\
\end{array}
$$
It follows from the existence of the classical $r$-matrix \cite{KLO} they are pairwise poisson commute.
Therefore,
they play the role of conservation laws of elliptic top
hierarchy on $\SLN$.
We have a tower of $\oh N(N+1)-1$ independent integrals of motion.
Note that $I_{k,k}=(\bfS^k)~(k=2,3\ldots,N)$.
 are the Casimir functions (\ref{fc}). By fixing their values we come  to the coadjoint orbit (\ref{orb}).
 In this way eventually we have $\oh N(N-1)$ integrals of motion. The number of integrals
 is equal to $\oh\dim\,\clO_\nu$ (\ref{dor}). Therefore, the elliptic top is completely integrable system.

\bigskip
\noindent
{\bf Extension of the top}\\
Consider the case 2. in (\ref{ca}).  Replace in expansion of the Lax operator (\ref{l1})
the coefficients $S_\al$ on $X_\al$
 \beq{l11}
L^{EET}(z)=\sum_{\al\in\ti\G_N}X_\al\varphi_{\al}(z)T_\al\,,
\eq
where
\beq{x}
\bfX=\sum_{\al\in\ti\G_N}X_\al T_\al\,.
\eq
We take here $\tr\,\bfX=X_0=0$ (see 1.(\ref{sc})). This condition corresponds to the $T^*\SLN$-bundle.
 To define the $T^*\SLN$-bundle one should in addition impose
 the constraint 2.(\ref{sc}), corresponding to the gauge fixing.
In these circumstances it is convenient to work with the not a completely gauged system.
In other words we will use in our constructions $\bfX^S$ (\ref{x}).

Using the expansion (\ref{lk}) we construct the tower of commuting integrals (\ref{ti}).
In terms of variables $X_c$ we have as above $\oh N(N-1)$ integrals of motion.
But this number is less then $\oh\dim\,T^*\clX^G$ (\ref{dcb}).
Remember that $\bfX^G$ is a special section of the bundle $T^*\clX^G$.
 The coordinates on $T^*\clX^G$ are $\clP$ and $\clQ$ (\ref{cav}).
 In these terms the quantities $I_{k,k}= ((\bfX^G)^k)$ are no longer Casimir functions with
 respect to the Darboux brackets (\ref{pb2}). With respect to these brackets we
 have $\oh N(N+1)-1$ integrals. This number is less then $\oh\dim\,T^*\clX^G$ (\ref{dcb}), but
 coincides with $\oh\dim\,T^*\clX^S$ (\ref{dcb1}).

In terms of the Darboux coordinates the Hamiltonian is
\beq{ha2}
H^{EET}=\oh\sum_{\al\in\ti{\Gamma}_N}X_\al\wp_\al X_{-\al}=
\oh\sum_{\al\in\ti{\Gamma}_N}(\clP\clQ)_\al\wp_\al (\clP\clQ)_{-\al}\,.
\eq

The Hamiltonian $H^{EET}$ is related to the Lax operator as in
(\ref{qh})
$$
((L^{EET})^2)(z)=I_{0,2}+I_{2,2}\wp(z)\,,~~I_{0,2}=-2H^{EET}\,.
$$

Let us write the equations of motion.
To define $(\clP\clQ)_\al$ (\ref{xr1}) explicitly  we introduce
 the basis $Y_a$ in the
the space of symmetric matrices $\gp^G$ (\ref{cd})
$$
Y_{a}= \oh(T_a+T^T_a)=
\oh(T_{a_1,a_2}+\bfe_N(a_1a_2)T_{a_1,-a_2})=
\frac{N}{4\pi\imath}\bfe_N(\frac{a_1a_2}2)Q^{a_1}(\La^{a_2}+\La^{-a_2})\,,~a\in\G_N\,.
$$
Then
$$
Y_aY_b=\sum_{\ga\in\ti\G_N}f_{a,b}^\ga T_{\ga}\,,
$$
where
\beq{str}
f_{a,b}^\ga=f_{a,b}\,,~(\ga=a+b)
\eq
 and
\beq{sco}
f_{a,b}=\frac{N}{2\pi\imath}\times
\Bigl(\bfe_N(a_1b_2-a_1b_2)
+\bfe_N(-a_1a_2-b_1b_2+a_2b_1-a_1b_2 )+
\Bigr.
\eq
$$
\Bigl.
\bfe_N(-a_1a_2-a_2b_1-a_1b_2)+
\bfe_N(b_1b_2+a_1b_2+a_2b_1)
\Bigr)
$$

There is the expabsion
\beq{p}
\clP=\sum_{a\in\G_N}p_aY_a\,,~~
\clQ=\sum_{b\in\G_N}q_bY_b\,.
\eq
\beq{sb}
~a_1=0,\ldots,N-1\,,~a_2=0,\ldots,N-1\,.
\eq

Since $\bfX=\sum_{\ga\in\ti\G_N}X_\ga T_\ga$, $X_\ga=(\clP\clQ)_{\ga}$
$$
X_{\ga}=\sum_{(m,n\in\G_N}\clP_m\clQ_nf_{m,n}\,,
$$
Because  $\tr\,\bfX=0$ there are the quadratic constraints
$$
\sum_{(m,-m\in\G_N}\clP_m\clQ_{-m}f_{m,-m}=0\,.
$$

The corresponding equations of motion are
$$
\p_t\clQ_a=\sum_b\clQ_b f_{a,b}\wp_{a+b}f_{-a,-b}\clP_{-m}\clQ_{-n}\,,
$$
$$
\p_t\clP_a=-\sum_b\clP_b f_{a,b}\wp_{a+b}f_{-a,-b}\clP_{-b}\clQ_{-a}\,.
$$


\subsection{Extension of Calogero-Mozer system}

\bigskip
\noindent
{\bf Elliptic Calopgero-Moser system with spin}\\
Consider  the Chevalley basis $\{e_j=E_{jj}\,,~~E_{jK}\}$ in the Lie algebra $\sln$.
The matrices
$\bfS=\sum S_{jk}E_{jk}$ define the spin variables and the diagonal matrices
$\bfu=\sum_ju_je_j$ and $\bfv=\sum_jv_je_j$ $(\sum\,u_j=\sum\,v_j=0)$
define the coordinates and momenta of N particles.
They have the canonical brackets $\{v_j,u_k\}=\de_{jk}$.
The coefficients $(S_{jj},S_{jk})$ satisfy the Poisson-Lie algebra $\sln$.
The polynomial
 $c_2=\tr\,\bfS^2$,...,$c_N=\tr\,\bfS^N$ (compare with (\ref{fc})) are
the Casimir functions with respect to these brackets. Fixing them one can invert the brackets and come to the non-degenerate form $\om^{KK}=(\bfS,g^{-1}Dg)$ on the orbit $\bfS=g^{-1}\nu g$ (\ref{kkf}).
This set forms coordinates on the phase space $\ti\clM^{CM}$
\beq{tm1}
\ti\clM^{CM}=\{\bfv\,,\,\bfu\,,\,\bfS\}\,.
\eq

There is residual gauge symmetry acting only on the spin variables
$H^\mC\in\SLN$:
\beq{ha}
g\to gh^{-1}\,,~~\bfS\to h\bfS h^{-1}\,.
\eq
The corresponding moment constraints are $S_{jj}=0$. To come to non-degenerate brackets one should
in addition fix the gauge,
for example as $S_{j,j+1}=1$.
The brackets for the variables $S_{jk}$ become the Dirac brackets corresponding to these constraints.
Thus we come to the phase space for the elliptic CM system with spin
\beq{pc}
\clM^{CM}=\ti\clM^{CM}//H^\mC=(\bfv,\bfu)|\,,~\{S_{jk}\,,\,j\neq k\,,\,S_{j,j+1}=1\}\,.
\eq
 It has dimension
\beq{dcm}
\dim_\mC\,\clM^{CM}=N(N-1)\,.
\eq
 Note that it coincides with $\dim\,\clO_\nu$ (\ref{dor}).

The Hamiltonian of the elliptic Calopgero-Moser system with spin has the form
\beq{hc}
H^{CM}=\oh(\bfv,\bfv)+\sum_{j\neq k}S_{jk}S_{kj}\wp(u_j-u_k)\,.
\eq



The  Lax operator  corresponding the  Calogero-Mozer has the quasi-periodicities
\beq{lo}
L^{CM}(z+1)=L^{CM}(z)\,,~~L^{CM}(z+\tau)=\clR L^{CM}(z)\clR^{-1}\,,
\eq
where $\clR=\di(\bfe(u_1),\ldots\bfe(u_N))$.
It follows from (\ref{A.14}) and (\ref{rk})  that
\beq{la0}
L^{CM}(z)=\sum_{j=1}^Nv_je_j+\sum_{j\neq k}S_{jk}\phi(u_{jk},z)E_{jk}\,,~~\,,~u_{jk}=u_j-u_k\,.
\eq
In terms of the Lax operator the Hamiltonian of the system (\ref{hc}) is defined as in (\ref{qh}).
In the similar way we obtain $\oh N(N-1)-1$ integrals of motion.

\bigskip
\noindent
{\bf Extension of the  system}\\
Assume now that the residue of the Lax operator
2.(\ref{ca}) is a section $\bfX$ (\ref{bx}) of the cotangent
bundle $T^*\clX^S$
$$
\bfX=\sum_{j=1}^NX_{jj}e_j+\sum_{j\neq k}X_{jk}E_{jk}\,.
$$
Then as in (\ref{la0}) the corresponding Lax operator $L^{ECM}$ of the extended Calogeroo-Moser model has the form
\beq{ela0}
L^{ECM}(z)=\sum_{j=1}^Nv_je_j+\sum_{j\neq k}X_{jk}\phi(u_{jk},z)E_{jk}\,,~~\,,~u_{jk}=u_j-u_k\,.
\eq
Here we assume that $X_{jj}=0$. These constraints come as above after the symplectic reduction
 by the diagonal matrices $H^\mC$.
The symplectic quotient is the phase space of the extended CM system (compare with
(\ref{tm1})
$$
\ti\clM^{ECM}=\{\bfv\,,\,\bfu\,,\,\bfX\}\,.
$$
Then
\beq{sqc}
\clM^{ECM}=\ti\clM^{ECM}//H^\mC=\{(\bfv\,,\,\bfu)\,,~X_{jk}\,(j\neq k)\,,\,X_{j,j+1}=1\}\,.
\eq
\beq{dec}
\dim\,\clM^{ECM}= N(N+1)-2
\eq
It coincides with $\dim_\mC\,T^*\clX^S$ (\ref{dcb1}).

The Hamiltonian of the system takes the form (see(\ref{hc})
$$
H^{ECM}=\oh(\bfv,\bfv)+\sum_{j\neq k}X_{jk}X_{kj}\wp(u_j-u_k)\,.
$$


As in the case of the elliptic top we use the expansion (\ref{lk}) to define the tower of commuting integrals (\ref{ti}).
We have as above $\oh N(N+1)-1$ integrals of motion.
 The quantities $I_{k,k}= ((\bfX)^k)$ are no longer Casimir functions with
 respect to the Darboux brackets (\ref{pb2}). With respect to these brackets we
 have $\oh N(N+1)-1$ integrals. This number is less then $\oh\dim\,T^*\clX^G$ (\ref{dcb}), but
 coincides with $\oh\dim\,T^*\clX^S$ (\ref{dcb1}).

Note that $H^\mC\subset\SLN$ lies in $\clX^S$
($H^\mC\notin\SON$).
 One can consider the additional left action of the subgroup $H^\mC$ on the moduli space $\clM^{ECM}$.
 The subgroup does not act on the variables $\bfv$ and $\bfu$ and acts on $g\to hg$ as in
(\ref{lda}). We have found that
$\clO_\nu=H\setminus\!\setminus_\nu T^*\clX^S$ (\ref{oga}). Since the left action does not touch
$(\bfv\,,\,\bfu)$ we obtain that
\beq{hec}
H^\mC\setminus\!\setminus_\nu \clM^{ECM}=
\clM^{ET}\sim\clO_\nu\,,~(\ref{pst})\,.
\eq

Describe the phase space of the extended CM system in terms of the
Darboux variables. Consider the basis
$$
Y_{jk}=\oh(E_{jk}+E_{kj})\,, \,e_j=E_{jj}
$$
in the space of symmetric matrices. As in (\ref{p})
\beq{qo}
\clP=\sum_jp_je_j+\sum_{j>k}p_{jk}Y_{jk}\,,~~
\clQ=\sum_jq_je_j+\sum_{j>k}q_{jk}Y_{jk}\,.
\eq
Then
$$
\bfX=\clP\clQ=\sum_aX_ae_a+\sum_{a\neq b}X_{ab}E_{ab}\,,
$$
Since
$$
e_lY_{jk}=f_{ljk}^{ab}E_{ab}\,,~~
f_{ljk}^{ab}=\oh (\de_{lj}\de_{la}\de_{kb}+\de_{lk}\de_{ka}\de_{jb})\,,
$$
$$
Y_{jk}e_l=f_{jkl}^{ab}E_{ab}\,,~~
f_{jkl}^{ab}=\oh (\de_{lk}\de_{ja}\de_{kb}+\de_{lj}\de_{ka}\de_{jb})\,,
$$
$$
Y_{jk}Y_{mn}=f_{jkmn}^{ab}E_{ab}\,,
$$
$$
f_{jkmn}^{ab}=
\f1{4}(\de_{aj}\de_{bn}\de_{km}+\de_{aj}\de_{bm}\de_{kn}+\de_{ak}\de_{bn}\de_{jm}+\de_{ak}\de_{bm}\de_{jn})\,.
$$
$$
f_{jkmn}^{aa}=
\f1{4}(\de_{aj}\de_{jn}\de_{km}+\de_{ak}\de_{km}\de_{kn}+
\de_{ak}\de_{kn}\de_{jm}+\de_{ka}\de_{km}\de_{jn})
$$
we obtain
\beq{qc}
X_a=p_aq_a +\sum_{jkmn}f_{jkmn}^{aa}p_{jk}q_{mn}\,,
\eq
$$
X_{ab}=\sum_{ljk}p_lq_{mn}f_{ljk}^{ab} +\sum_{jkmn}f_{jkmn}^{ab}p_{jk}q_{mn}\,.
$$

The Hamiltonian $H^{ECM}$ (\ref{hec}) is the quartic polynomial in terms of
the Darboux variables. In addition, there are the quadratic constraints $X_a=0$ (\ref{qc}).

\vspace{0.3cm}
{\small {\bf Acknowledgments.}  The work was supported by RFBR grant 19-51-18006-BOLG
 }


\end{document}